\documentclass{article}
\usepackage{amsmath,amssymb,yhmath,bm}
\usepackage{pstricks,pst-node,pst-text,pst-3d, psfrag,graphicx}
\usepackage{slashed} 
\usepackage{color}
 
\newcommand {\be}  {\begin{equation}}  
\newcommand {\ee}  {\end{equation}}

\title{1974: the discovery of the first binary pulsar}
\author{Thibault {\sc Damour}$^1$}
\date{{\small E-mail: damour@ihes.fr} \\ { \ } \\
{\small$^1$ Institut des Hautes Etudes Scientifiques, 35, route de Chartres, \\ 91440 Bures-sur-Yvette, France}}

\begin{document}

\maketitle

\begin{abstract}
The 1974 discovery, by Russell A. Hulse and Joseph H. Taylor, of the first binary pulsar PSR B1913+16, opened up new possibilities for the study of relativistic gravity. PSR B1913+16, as well as several other binary pulsars, provided {\it direct} observational proofs that gravity propagates at the velocity of light and has a quadrupolar structure. Binary pulsars also provided accurate tests of the strong-field regime of relativistic gravity. General Relativity has passed all the binary pulsar tests with flying colors. The discovery of binary pulsars had also very important consequences for astrophysics: accurate measurement of neutron star masses, improved understanding of the possible evolution scenarios for the co-evolution of binary stars, proof of the existence of binary neutron stars emitting gravitational waves for hundreds of millions of years, before coalescing in catastrophic events radiating intense gravitational-wave signals, and probably leading also to  important emissions of electromagnetic radiation and neutrinos. This article reviews the history of the discovery of the first binary pulsar, and describes both its immediate impact, and its longer-term effect on theoretical and experimental studies of relativistic gravity.
\end{abstract}

\vglue 1cm 

\noindent PACS numbers: 04.20.-q, 04.25.Nx, 04.30.-w, 04.80.Cc, 97.60.Gb

\newpage

\section{Prelude}

The first binary pulsar, PSR 1913+16 (henceforth referred to by its currently recommended name, PSR B1913+16), was discovered, at the Arecibo radio telescope, on July 2, 1974 in the middle of a systematic pulsar search carried out by Russell A. Hulse as a basis for his PhD thesis under the guidance of Joseph H. Taylor \cite{Hulse:1994zz}. The research proposal behind this collaborative work of Hulse and Taylor had been submitted by Joe Taylor to the US National Science Foundation in September 1972 and had mentionned, among the motivations for planning an extensive pulsar survey, that it would be highly desirable ``$\ldots$ to find even {\it one} example of a pulsar in a binary system, for measurement of its parameters could yield the pulsar mass, an extremely important number'' \cite{Taylor:1994zz}. The discovery of a binary pulsar was not, however, the main motivation for the new pulsar search undertaken by Joe Taylor and Russell Hulse. At the time of Joe's proposal (September 1972),  sixty seven pulsars had been discovered since the discovery of the first pulsar by Jocelyn Bell and Antony Hewish at Cambridge University (UK) in 1967. However, Joe Taylor had realized that there was room for a systematic, high-sensitivity pulsar survey using the largest existing radio telescope, and, most importantly, using a computerized search, based on an efficient dispersion-compensating algorithm, in the three-dimensional space spanned by dispersion measure, period and pulse width. This computerized search algorithm, combined with the large Arecibo telescope, achieved a pulsar detection sensitivity over ten times better than that reached by any previous pulsar search \cite{HulseTaylor1974}. This high-sensitivity was crucial for the discovery of the first binary pulsar. Indeed, PSR B1913+16 is a rather weak pulsar, and its initial detection occurred at $7.25 \, \sigma$, i.e. just above the search detection threshold of $7.0 \, \sigma$. [Such a seemingly high detection threshold was necessary to keep the false detection rate reasonably low in spite of the large parameter space being searched.] In other words, the initial discovery of the first binary pulsar was a very close call, but the success was due to all the effort put by Russell Hulse and Joe Taylor on getting every last bit of possible sensitivity out of their computerized pulsar search algorithm.

\smallskip

When Russell Hulse discovered PSR B1913+16 on July 2, 1974, he inscribed the comment ``fantastic!'' on the discovery form because its 59 msec period made it the second fastest pulsar known at the time (after the famous 33 msec Crab nebula pulsar). The binary nature of PSR B1913+16 (which made it a really fantastic pulsar) was discovered by Hulse only about two months later, in early September through a careful and dedicated study of its (pulsing) period. Indeed, the first attempt (on August 25, 1974) to obtain a more accurate value of the pulsar period was frustrated by a perplexing result: after correcting for the effect of the Doppler shift due to the Earth' motion, the pulsar periods determined during separate, short observations differed by up to 80 microseconds from day to day, i.e. by an enormous amount compared to normal (isolated) pulsars. After several weeks of detective work (using a special computerized de-dispersing analysis) Russell Hulse contacted on September 18, from Arecibo, Joe Taylor at the University of Massachusetts in Amherst to tell him the amazing news behind these large apparent variations in pulsar period, namely that the pulsar was in a high-velocity binary orbit with about an 8 hour period. Soon Joe Taylor was on a plane to Arecibo. He arrived with a special hardware de-disperser which allowed a more efficient study of the apparent period variations, soon leading to an accurate determination of the complete ``velocity curve'' of PSR B1913+16, i.e. of the variation with orbital phase of the radial velocity of the pulsar on its orbit. Within a month they were ready to report the discovery of the first pulsar in a binary system.

\section{The discovery paper}
Their discovery paper \cite{Hulse:1974eb} (which was received on October 18, 1974) features an accurate determination of the velocity curve of PSR B1913+16, from which (using a least-squares fit) they inferred the orbital period\footnote{Here, and below, the parentheses indicate the error on the corresponding last digit(s).}, $P_b = 27908(7)s$, the projected semi-major axis of the pulsar orbit, $a_1 \sin i = 1.00(2) \, R_{\odot}$, its eccentricity $e = 0.615(10)$, the longitude of the periastron\footnote{As was noted at the time, the discovery values of both $e$ and $\omega$ are close to some special numbers: $e$ to the golden ratio $\frac12 (\sqrt5 - 1) \simeq 0.618$ and $\omega_0$ to $180$ degrees $= \pi$.}, $\omega_0 = 179(1)$ degree, the radial velocity semiamplitude $K_1 = 199(5)$ kms$^{-1}$ and the mass function,
$$
f(m_1,m_2) \equiv (m_2 \sin i)^3 / (m_1 + m_2)^2 = 0.13(1) \, M_{\odot} \, .
$$

As they did not know (at the time) the inclination angle $i$, their measurement of the mass function was compatible with a wide range of values for the mass, $m_1$, of the pulsar and the mass, $m_2$, of its companion. However, by restricting their attention to values of $m_1$ ``thought to be reasonable for neutron stars'' (namely $0.3 \leq m_1 / M_{\odot} \leq 1.5$), and by using the fact that no eclipses were observed, they (``virtually'') ruled out the possibility that the companion be a main-sequence star, and they concluded that ``the companion must be a compact object, probably a neutron star or a black hole''; adding that ``a white-dwarf companion cannot be ruled out, but seems unlikely for evolutionary reasons''.

\smallskip

In their ``additional observations'' they mentionned several possibilities that turned out to be important for future developments. First, they mentionned the ``exciting possibility'' that the unseen companion be also a radiofrequency pulsar, so that the system would be a ````double-line'' spectroscopic binary''. Second, they indicate that much more accurate timing data can be obtained by recording the absolute time of arrival of the pulses, and that they started acquiring such time-of-arrival (TOA) data, adding that:

\begin{quote}
``This $\ldots$ will allow a number of interesting gravitational and relativistic phenomena to be studied. The binary configuration provides a nearly ideal relativity laboratory including an accurate clock in a high-speed, eccentric orbit, and a strong gravitational field. We note, for example, that the changes of both $v^2/c^2$ and $GM/c^2 r$ during the orbit are sufficient to cause changes in observed period of several parts in $10^6$. Therefore, both the relativistic Doppler shift and the gravitational redshift will be easily measurable. Furthermore, the general-relativistic advance of periastron should amount to about $4^{\rm o}$ per year, which will be detectable in a short time. The measurements of these effects, not usually observable in spectroscopic binaries, would allow the orbit inclination and the individual masses to be obtained.''
\end{quote}

As last good news, they remarked that PSR B1913+16 promised to be a {\it clean} system, indeed: ``No changes in dispersion measure exceeding $\pm \, 20$ cm$^{-3}$ pc [on a total $DM = 167(5)$ cm$^{-3}$ pc] have been observed over the binary period, so it is clear that at most a small fraction of the dispersion can arise from electrons within the binary orbit.''

\section{Immediate impact of the discovery}

The announcement\footnote{Before submitting, and sending around in preprint form, their October 18 discovery paper, Joe Taylor and Russell Hulse announced their discovery in the October 4, 1974 IAU circular n$^{\rm o}$~2704 as a short telegram. Moreover, Joe Taylor gave a colloquium on the binary pulsar at Cornell University on October 2, and
gave further talks at Stanford, Princeton and Harvard during the week of October 14-18. [Joe also remembers hand-delivering the manuscript of their discovery paper 
on October 18 to Alex Dalgarno, at the ApJ Letters office, at the Harvard College Observatory.]}, in October 1974, of the discovery created a lot of excitement worldwide. The author (who had recently arrived at Princeton University as a 23-year old postdoc) vividly remembers hearing about the discovery at one of the weekly (Tuesday) lunches organized by John Bahcall at the Institute for Advanced Study. These lunches, whose role was to circulate news and exchange information about recent advances or discoveries in astrophysics, were well attended by the whole Princeton physics and astrophysics community (to which notably belonged Remo Ruffini, who had taken the author to this memorable lunch). Excitement filled the air when the discovery was announced at this lunch.

\smallskip

In the following months, a flurry of papers discussed various potential physical or astrophysical consequences of the Hulse-Taylor discovery, notably Refs.~\cite{Wagoner1974,Will1974,Brecher1974,MastersRoberts1974,EspositoHarrison1974,Eardley1974,Brumberg1974,Flannery1974,CraigWheeler1974,DamourRuffini1974,BarkerOConnell1975,BlandfordTeukolsky1975,RobertsMastersArnett1975}. Let us only highlight here some of the most significant suggestions made in these papers. Ref.~\cite{Wagoner1974} pointed out that the observation of the secular decrease of the orbital period of PSR B1913+16, $\dot P_b \equiv dP_b / dt$, would constitute a ``test for the existence of gravitational radiation''. Ref.~\cite{Eardley1974} pointed out that the orbital period decay $\dot P_b$ was a sensitive test of alternative relativistic theories of gravity, and notably of tensor-scalar theories (such as the Jordan-Brans-Dicke theory) because the emission of dipole gravitation radiation in theories containing scalar excitations would generally be much larger than the usual quadrupolar emission. Ref.~\cite{DamourRuffini1974}\footnote{Note that this paper is the only one which was published in 1974, i.e. even before the Hulse-Taylor discovery paper which was published in January 1975. This was due to the celerity of publication in the comptes rendus of the French academy of sciences.} (see also \cite{EspositoHarrison1974}) pointed out that the relativistic spin-orbit precession of the pulsar spin axis (of a few degrees per year) will allow not only to test General Relativity, but also to observe, for the first time, the pulsar emission process at varying angles. It also mentionned that the pulsar emission might periodically become invisible from Earth because of this slow spin-orbit precession. [Ref.~\cite{DamourRuffini1974} estimated the spin-orbit precession frequency by using the extreme-mass-ratio limit $m_1 \ll m_2$; the more general, comparable-mass expression was provided a few months later, see Refs.~\cite{BarkerOConnell1975,Damour1975,BoernerEhlersRudolph1975,Cho:1975yr}.] 
A detailed analysis of the effect of spin-orbit precession on the observed pulse width and polarisation sweep was  presented in \cite{HariDassRadhakrishnan1975}.
From the astrophysical point of view, the study of possible consistent evolutionary scenarios leading to the formation of binary pulsars was pioneered in Ref.~\cite{Flannery1974}.

\section{Relativistic timing of binary pulsars}

After the initial flurry of papers spurred by the announcement of the discovery of the first binary pulsar, the rate of production of papers dealing with theoretical aspects of the dynamics and timing of binary systems soon dropped down. However, a small sequence of papers dealing with the relativistic theory of the timing of binary pulsars had a significant impact on the later data analysis of binary pulsars. This line of work was pioneered by a work of R. Blandford and S.A. Teukolsky \cite{Blandford1976} which derived the first version of the ``timing formula'' of a binary pulsar, i.e. a mathematical expression giving the functional relation between the integer $N$ labelling the $N$th pulse and its time of arrival (TOA) at the Earth as measured by an observer on the Earth. The use of such timing formulas to analyze the observed sequence of TOA's proved to be instrumental for extracting accurate physical information from binary pulsars. The timing formula derived in \cite{Blandford1976} included only some of the relativistic effects in binary pulsars, namely: (i) the combined second-order Doppler shift and gravitational redshift (measured by a parameter called $\gamma$); (ii) the secular advance of the periastron $(\dot\omega)$; (iii) the secular change of the orbital period $(\dot P_b)$; and (iv) the relativistic time delay linked to the propagation of the pulsar signal in the gravitational field of the companion $(m_2 ; \sin i)$. A subsequent work \cite{Smarr1976} included the aberration effect linked to the orbital velocity of the pulsar.

\smallskip

The timing effects linked to the (periodic) first post-Newtonian (1PN) contributions to the two-body dynamics (i.e. the $O(v^2/c^2)$ Einsteinian corrections to Newton's $1/r^2$ law) were tackled in later papers \cite{Epstein1977}, \cite{Haugan1985}, \cite{DD86}. The post-Newtonian-accurate timing formulas of Refs.~\cite{Epstein1977} and \cite{Haugan1985} had several unsatisfactory features that were corrected in the ``DD'' timing formula \cite {DD86}, which was based on a new, simplified ``quasi-Keplerian'' representation of the 1PN-accurate motion \cite{DD85}. See \cite{Weisberg:1984zz}  for an early use of the Epstein-Haugan timing model, and 
\cite{Taylor:1989sw} for a comparison of the performances of the various timing models.

\smallskip

The DD timing formula later led to defining the ``parametrized post-Keplerian'' (PPK) formalism \cite{Damour:1987wn}, \cite{Damour:1991rd}, a strong-field analogue of the parametrized post-Newtonian (PPN) formalism \cite{Will:2014xja} designed to extract dynamical information from binary-pulsar data in a phenomenological, theory-independent way. The use of this formalism has allowed one to test strong-field aspects of relativistic gravitation (see below).

\section{Establishing that gravity propagates as predicted by General Relativity}

One of the first, and most spectacular, legacies of the discovery of binary pulsars has been to provide direct experimental evidence for the reality of gravitational radiation damping in binary systems. This was first publicly announced by Joe Taylor in December 1978 at the 9$^{\rm th}$ Texas Symposium (Munich, Germany) and was soon published \cite{Taylor:1979zz}. The experimental evidence consists of observing a secular decrease of the orbital period (i.e. a negative value of $\dot P_b$). The observed value $\dot P_b^{\rm obs}$ is then compared to the value of $\dot P_b$ predicted by Einstein's theory of General Relativity as a function of two other pulsar-timing observables, namely the secular rate of periastron advance $\dot\omega$ and the combined second-order-Doppler-gravitational-redshift timing parameter $\gamma$. In this first announcement the ratio between the observed and the GR-predicted value of $\dot P_b$ was
\begin{equation}
\label{eq5.1}
\left[\frac{\dot P_b^{\rm obs}}{\dot P_b^{\rm GR} \left[\dot\omega^{\rm obs} , \gamma^{\rm obs}\right]}\right]_{1913+16} = 1.3 \pm 0.3 \, .
\end{equation}
This remarkable observational discovery spurred a lot of theoretical work on gravitational-radiation-related effects in binary systems. Even before the announcement some authors had pointed out some of the shortcomings of the derivation of both gravitational radiation damping and gravitational radiation energy loss in binary systems \cite{Ehlers:1976ji}, \cite{Rosenblum:1978zr}. After the announcement, many authors worked towards improving the theoretical understanding of gravitational radiation effects in binary systems \cite{Schutz:1980xm}, \cite{Kates:1980kr}, \cite{Walker:1980tm}, \cite{Papapetrou:1981me}, \cite{Rosenblum:1981kt}, \cite{Bel:1981be}, \cite{Damour:1981bh}, \cite{Damour:1982wm}, \cite{Damour:1983tz}. Note that the last two references provided a direct computation of the observed orbital-period derivative $\dot P_b$ from general relativistic two-body equations of motion valid for strongly self-gravitating bodies and complete up to fractional corrections of order $(v/c)^5$. This computation (see also \cite{Schafer1985}, \cite{Kopeikin1985}) shows that the observation of $\dot P_b$ in the TOAs of a binary pulsar is a {\it direct} effect of the retarded propagation (at the speed of light, and with a quadrupolar structure) of the gravitational interaction between the companion and the pulsar. In that sense, the Hulse-Taylor pulsar provides a {\it direct} observational proof that gravity propagates at the speed of light, and has a quadrupolar structure.

\smallskip

The latter point is confirmed by the theoretical computation of $\dot P_b$ in alternative theories of gravity where the non purely quadrupolar (i.e. non purely spin 2) structure of the gravitational interaction generically induces drastic changes in the theoretically predicted function $\dot P_b^{\rm theory} \left[\dot\omega , \gamma \right]$ \cite{Eardley1974}, \cite{Will:1977zz}, \cite{Will:1989sk}, \cite{Damour:1992we}. Such drastic changes are generally incompatible with the observed  $\dot P_b$~\cite{Weisberg:1981bh}.

\smallskip

On the observational front, the initial result, Eq. (\ref{eq5.1}), was later confirmed and refined~\cite{Taylor:1982zz,Taylor:1989sw}. 
When, the observational precision on $\dot P_b^{\rm obs}$ got better than 1\%, it became necessary to take into account non-GR contributions to $\dot P_b^{\rm obs}$ coming from the galactic accelerations of the pulsar and the Sun, and from the proper motion of the  pulsar \cite{Damour:1990wz}. After taking these contributions into account the ratio $\dot P_b^{{\rm obs} - {\rm gal}} / \dot P_b^{\rm GR}$ yielded a 0.8\% confirmation of General Relativity. The current timing data \cite{Weisberg:2010zz} yield.
\begin{equation}
\label{eq5.2}
\left[\frac{\dot P_b^{\rm obs} - \dot P_b^{\rm gal}}{\dot P_b^{\rm GR} \left[\dot\omega^{\rm obs} , \gamma^{\rm obs}\right]}\right]_{1913+16} = 0.997 \pm 0.002 \, ,
\end{equation}
i.e. an experimental evidence for the reality of gravitational radiation damping forces at the $(-3 \pm 2) \times 10^{-3}$ level. Several other binary pulsars have given similar confirmations of the reality of gravitational radiation (for reviews of pulsar tests see \cite{Will:2014xja}, and chapter 21 in the Review of Particle Physics \cite{Agashe:2014kda}).

\smallskip

Let us note in particular the stringent tests of the quadrupolar (rather than dipolar) structure of gravitational radiation damping obtained from the measurement of the orbital period decay $\dot P_b$ of several dissymetric pulsar-white dwarf binary systems. The most stringent constraint is obtained from the low-eccentricity 8.5 hour pulsar-white dwarf system PSR J1738+0333 \cite{Freire:2012mg}.

\section{Testing strong-field gravity}
\setcounter{equation}{0}

Another very important legacy of the 1974 discovery has been to provide clean and accurate tests of the strong-field regime of relativistic gravity. Indeed, contrary to the solar system (which is made up of weakly self-gravitating bodies, so that the curved spacetime metric $g_{\mu\nu} (x^{\lambda})$ is everywhere a small deformation of a flat metric $\eta_{\mu\nu}$) a binary pulsar contains at least one strongly self-gravitating body, namely the rotating neutron star emitting the observed radio pulses, so that the metric $g_{\mu\nu} (x^{\lambda})$ near it significantly differs from $\eta_{\mu\nu}$. Indeed, the surface value of, say, $h_{00} (R) \equiv \eta_{00} - g_{00} \simeq 2Gm/c^2R$ is of order $0.4$ for a neutron star, i.e. larger by a factor $\sim 10^8$ than the surface potential of the Earth, and a mere factor $2.5$ below the black hole limit $(h_{00} (R) = 1)$. The large deviation of the spacetime metric away from its Minkowski value opens the possibility of testing the strong-field regime of relativistic gravity, separately from its radiative aspects. [The $\dot P_b - \dot \omega - \gamma$ test discussed in the previous section mixes radiative effects (in $\dot P_b$) with static strong-field effects (in $\dot\omega$ and $\gamma$).] A general phenomenological formalism for using binary pulsar data to test strong-field gravity was presented in \cite{Damour:1991rd}. It is called the ``parametrized post-Keplerian'' (PPK) formalism. It was shown in Ref.~\cite{Damour:1991rd} that, in principle, as many as fifteen tests of relativistic gravity can be extracted from the observational data of a single binary pulsar. The data used for this purpose are both timing data and pulse-structure data. The phenomenological analysis of pulsar timing data is done by using the general, theory-independent version of the DD timing formula mentioned in Section~4.

\smallskip

The first experimental constraints on strong-field relativistic gravity were obtained in Ref.~\cite{Taylor:1993zz}. They used 10 years of timing data of PSR B1913+16, and one year of timing data of the second discovered (double-neutron-star) binary pulsar PSR B1534+12 \cite{Wolszczan:1991kj}. In particular, the PSR B1534+12 timing data allowed one to separately measure 4 phenomenological parameters: $\dot\omega^{\rm obs}$, $\gamma^{\rm obs}$, $r^{\rm obs}$ and $s^{\rm obs}$, where $r$ measures the ``range'' and $s$ the ``shape'' of the Shapiro time delay caused by the companion. In any given theory of relativistic gravity the 4 phenomenological parameters $\dot\omega^{\rm obs}$, $\gamma^{\rm obs}$, $r^{\rm obs}$, $s^{\rm obs}$ can be expressed as (theory-dependent) functions of the mass $m_1$ of the pulsar, the mass $m_2$ of its companion, and of the Keplerian parameters. One can then use (for a given theory) the two equations $\dot\omega^{\rm obs} = \dot\omega^{\rm thy} [m_1,m_2]$, $\gamma^{\rm obs} = \gamma^{\rm thy} [m_1,m_2]$ to determine $m_1$ and $m_2$ in terms of $\dot\omega^{\rm obs}$ and $\gamma^{\rm obs}$. This leads to two theory-dependent predictions for $r=r^{\rm thy} [\dot\omega , \gamma]$ and $s = s^{\rm thy} [\dot\omega , \gamma]$. Comparing the two theory-dependent predictions $r^{\rm thy} \left[ \dot\omega^{\rm obs} , \gamma^{\rm obs} \right]$ and $s^{\rm thy} \left[ \dot\omega^{\rm obs} , \gamma^{\rm obs} \right]$ to the two corresponding observational values $r^{\rm obs}$, $s^{\rm obs}$ then leads to {\it two} tests of strong-field gravity. Ref.~\cite{Taylor:1993zz} found that General Relativity passes these two strong-field tests. For instance, one had already at the time (i.e. when using only one year of PSR B1534+12 timing data)
\begin{equation}
\label{eq6.1}
\left[\frac{s^{\rm obs}}{s^{\rm GR} \left[ \dot\omega^{\rm obs} , \gamma^{\rm obs}\right]}\right]_{1534+12} = \frac{0.986 (7)}{0.982} = 1.004(7) \, .
\end{equation}

\smallskip

Theoretical investigations of the strong-field predictions of alternative relativistic theories of gravity \cite{Eardley1974,Will:1977zz,Will:1993ns,Damour:1992we,Damour:1993hw,Damour:1996ke,Damour:1998jk,EspositoFarese:2004tu,EspositoFarese:2009ta,Mirshekari:2013vb} have explicitly shown that while strong self-gravity effects are ``effaced'' in GR (in the sense that they can be renormalized away by including them in the definition of the two masses $m_1 , m_2$), this is not so in most alternative theories of gravity. It is generically found that the theoretical predictions relating PPK observables to the two masses, $\dot\omega^{\rm thy} [m_1,m_2]$, $\gamma^{\rm thy} [m_1 , m_2] , \ldots$, are significantly modified by self-gravity effects in alternative theories of gravity. As a consequence theoretically-predicted relations between PPK observables, such as the function $s^{\rm thy} [\dot\omega , \gamma]$, are modified by self-gravity effects. This explicitly shows that a binary-pulsar test of the type of Eq.~(\ref{eq6.1}) does indeed constrain the strong-field regime of relativistic gravity. Early examples of such effects \cite{Will:1977zz,Damour:1992we} were derived within ill-motivated theories, containing negative-energy excitations. It was later found that non-perturbative strong-field effects can however develop within certain tensor-scalar theories, even in cases where the weak-field limit of these theories is arbitrarily close to GR \cite{Damour:1993hw}.

\smallskip

Accurate non radiative strong-field tests have been obtained in three binary pulsar systems: PSR B1534+12 \cite{Fonseca:2014qla} (two tests), PSR J1141-6545 \cite{Bhat:2008ck} using $\sin i$ from \cite{Ord:2002bu} (one $s - \dot\omega - \gamma$ test) and PSR J0737-3039 \cite{Kramer:2006nb} (four tests). [For a review and more references see \cite{Agashe:2014kda}.] The last cited binary pulsar is the remarkable {\it double} pulsar \cite{Burgay:2003jj,Lyne:2004cj}, made of two neutron stars which are both observable as pulsars. This system has given five independent tests of relativistic gravity \cite{Kramer:2006nb,Breton:2008xy,Kramer:2009zza}.

\smallskip

General Relativity passes all the current strong-field tests with flying colors. The most accurate strong-field confirmation of GR is at the $5 \times 10^{-4}$ level. It was obtained from double binary pulsar data combining the observables $s,\dot\omega$ and the ratio $R = x_B / x_A$ between the projected semi-major axes of the two pulsars~\cite{Kramer:2006nb}:
\begin{equation}
\label{eq6.2}
\left[\frac{s^{\rm obs}}{s^{\rm GR} \left[ \dot\omega^{\rm obs} , R^{\rm obs}\right]}\right]_{0737-3039} = 1.0000(5) \, .
\end{equation}

The (radiative or nonradiative) constraints on tensor-scalar theories provided by the various binary pulsar ``experiments'' (involving double neutron-star or neutron-star white dwarf systems) have been analyzed \cite{Damour:1996ke,Damour:1998jk,EspositoFarese:2004cc,Freire:2012mg} and shown to exclude a large portion of the parameter space allowed by solar-system tests.

\smallskip

Finally, measurements over several years of the pulse profiles of various pulsars have detected secular profile changes compatible with the prediction \cite{DamourRuffini1974} that the general relativistic spin-orbit coupling should cause a secular change in the orientation of the pulsar beam with respect to the line of sight (``geodetic precession''). Such confirmations of general-relativistic spin-orbit effects were obtained in PSR B1913+16 \cite{Weisberg:1989}, \cite{Kramer:1998id}, \cite{WeisbergTaylor2002}, PSR B1534+12 \cite{Stairs:2004ye}, \cite{Fonseca:2014qla}, PSR J1141-6545 \cite{Hotan:2004ub},  PSR J0737-3039 \cite{Breton:2008xy}, and PSR J1906+0746 \cite{vanLeeuwen:2014sca}. Note that while one has actual measurements of  geodetic precession rates for PSR B1534+12  and  PSR J0737-3039, the evidence in the
other systems is purely qualitative.
In this respect, it is interesting to note (as Joe Taylor did in some talks) that (according to the precession model of \cite{Kramer:1998id}) PSR B1913+16 became visible from the Earth around 1941 (i.e. around the birth of Joe himself) and will become invisible around 2025. Then it will disappear for about 200 years before reappearing again as a pulsar visible from the Earth. It seems that other binary pulsars have similarly appeared or disappeared over a relatively small number of years.

\smallskip

Let us finally mention the remarkable recent  discovery of a millisecond pulsar in a {\it triple} system \cite{Ransom:2014xla}: PSR J0337+1715 is a 2.73 millisecond pulsar, with very good timing precision, around which {\it two} white dwarfs orbit in hierarchical, coplanar, nearly circular orbits. The gravitational field of the outer white dwarf (with orbital period 327.26 days) accelerates the two members (pulsar and white dwarf) of the inner binary (with orbital period 1.629 days). This system has the potential to test the strong equivalence principle (SEP) for strongly self-gravitating bodies (i.e. the universality of free fall of self-gravitating bodies) to an unprecedented level of accuracy. Data from several nearly circular  binary systems (made of a neutron star and a white dwarf) have already led to strong-field confirmations of the SEP (at the $4.6 \times 10^{-3}$ level), by checking  that a strongly self-gravitating neutron star and a relatively weakly self-gravitating white dwarf fall with the same acceleration in the gravitational field of the Galaxy \cite{Damour:1991rq}, \cite{Freire:2012nb}. In the triple system PSR J0337+1715 the perturbing acceleration field (due to the outer white dwarf) is at least 6 orders of magnitude larger than the Galactic acceleration used in previous pulsar tests of the SEP. This opens the promise of an extremely interesting new test of strong-field gravity.

\section{Binary pulsars and astrophysics}
\setcounter{equation}{0}

Up to here we focused on the legacies of the 1974 discovery for relativistic gravity. However, this discovery had also very important consequences for astrophysics. Let us indicate some of them.

\smallskip

First, it led to the first accurate measurement of neutron star masses. For instance, the masses of the pulsar and its companion in PSR B1913+16 were found (when assuming a GR-based timing model \cite{Taylor:1989sw}) to be \cite{Weisberg:2010zz}
\begin{equation}
\label{eq7.1}
m_1 = 1.4398(2) \, M_{\odot} \ ; \quad m_2 = 1.3886(2) \, M_{\odot} \, .
\end{equation}

For a compilation of observed neutron star masses (obtained either through pulsar timing or other methods) and references see \cite{Lattimer:2012nd}. Note that, recently, pulsar masses close to $2M_{\odot}$ have been discovered \cite{Demorest:2010bx,Antoniadis:2013pzd}. These large neutron star masses put strong constraints on the equation of state of nuclear matter \cite{Hebeler:2013nza}.

\smallskip

The discovery of binary pulsars, as well as the later discovery of fast-spinning ``millisecond'' isolated pulsars \cite{Backer1982}, has given an enormous impetus to the development of astrophysical scenarios for the co-evolution of binary stars. These studies had been initiated in 1972 with the discovery, by the UHURU satellite, of pulsating X-ray sources in close binaries around main-sequence stars \cite{Schreier1972}. In April 1973 (and therefore before the discovery of the first binary pulsar) G.S. Bisnovatyi-Koyan and B.V. Komberg had already suggested that some pulsating X-ray sources in binaries may later in their life, after their massive companion stars have exploded as a supernova, become observable as binary radio pulsars \cite{Bisnovatyi-Kogan1974}. For an authorative review of the formation and evolution of relativistic binaries see the contribution of E. P. J. van den Heuvel to the book \cite{ColpiCasella2009}.

\smallskip

Another legacy of the 1974 discovery concerns the later life of binary neutron stars. Radio pulsars are thought of having a limited lifetime (as active pulsars). But the discovery of binary pulsars shows the existence of a population of binary neutron stars that will emit gravitational waves for hundreds of millions of years. For instance, PSR B1913+16 will coalesce, because of gravitational radiation damping, in a few hundred million years. In the last few minutes before coalescence, such a binary neutron star will be a strong emitter of gravitational radiation (as was first pointed out by Freeman Dyson \cite{Dyson1963}). These sources are some of the prime targets for the upcoming network of advanced ground-based interferometric gravitational wave detectors. Binary pulsar observations are crucially used to estimate the number of gravitational-wave signals expected in such detectors (see \cite{Abadie:2010cf} and references therein). One expects to extract interesting scientific information from the future observation of inspiralling and coalescing binary neutron stars. In particular, the gravitational wave signal from the late inspiral will give useful constraints on the equation of state of nuclear matter \cite{Damour:2012yf,Lackey:2013axa,DelPozzo:2013ala}. In addition, it is hoped that the final coalescence process will give rise to catastrophic events leading to an important emission of electromagnetic radiation and neutrinos. These events might be related to the so-called {\it short} gamma-ray bursts \cite{Rezzolla:2011da}.

\section{Concluding remarks}

The 1974 discovery of the first binary pulsar has given us a cornucopia of important scientific benefits. The most spectacular ones concern the first experimental evidence that Einstein's theory of General Relativity is valid beyond the usually tested quasi-stationary, weak-field regime. Indeed, binary pulsar data have probed, for the first time, relativistic gravity in regimes involving (either together or separately) radiative effects and strong-field effects. The citation accompanying the award, in October 1993, of the Nobel Prize in Physics to Russell A. Hulse and Joseph H. Taylor read: ``for their discovery of a new type of pulsar, a discovery that has opened up new possibilities for the study of gravitation''. As we have discussed, these new possibilities for studying gravitation have been even more sucessful than what was envisaged in the months following the discovery.

\smallskip

Even more importantly, the class of systems discovered by Hulse and Taylor promises to bring new discoveries in the near future, through the physics of the late stages of evolution of compact binaries: gravitational waves, probes of nuclear-matter equation of state, possible connection with gamma-ray bursts,$\ldots$ Let us finally mention the hope that radio pulsars in orbit around a black hole will soon be discovered. The black hole companion could be either a $\sim 10 \, M_{\odot}$ black hole, or, possibly, a much more massive black hole. Recently, a magnetar was discovered near the massive ($\sim 4 \times 10^6 \, M_{\odot}$) black hole at the center of our Galaxy \cite{Mori:2013yda}. Searches are underway for discovering pulsars having better timing stability, and closer to the galactic center. Such a discovery would be a fantastic new  milestone for General Relativity.

\smallskip

\noindent  {\bf Acknowledgments}  I am grateful to Joe Taylor for a careful reading of the manuscript, and for additional historical information.

\end{document}